\documentclass[fleqn,usenatbib,useAMS]{mnras}

\usepackage{graphicx}	
\usepackage{amsmath}	
\usepackage{multicol}        
\usepackage{bm}		
\usepackage{pdflscape}	
\usepackage{orcidlink}
\usepackage{xspace}
\usepackage{gensymb}
\usepackage{caption}
\usepackage{float}

\newcommand{\mj}{$M_{\rm J}$}
\newcommand\exofast{$\texttt{EXOFASTv2}$\xspace}

\newcommand{\thislambda}{$|\lambda| = 12.7 \pm 1.3 \degree$}

\newcommand{\msu}{Center for Data Intensive and Time Domain Astronomy, Department of Physics and Astronomy, Michigan State University, East Lansing, MI 48824, USA}
\newcommand{\usq}{Centre for Astrophysics, University of Southern Queensland, West Street, Toowoomba, QLD 4350, Australia}
\newcommand{\cfa}{Center for Astrophysics \textbar \ Harvard \& Smithsonian, 60 Garden St, Cambridge, MA 02138, USA}

\usepackage[T1]{fontenc}
\usepackage{ae,aecompl}

\usepackage{newtxtext,newtxmath}

\title[OATMEAL III]{The OATMEAL Survey. III. An Aligned Transiting Warm Brown Dwarf and Evidence for Quiescent Brown Dwarf Migration\thanks{This paper includes data gathered with the 6.5 meter Magellan Telescopes located at Las Campanas Observatory, Chile.}}

\author[N. Vowell et al.]{
Noah Vowell$^{1}$\thanks{E-mail: vowellno@msu.edu}\orcidlink{0000-0002-0701-4005},
Jiayin Dong$^{2}$\orcidlink{0000-0002-3610-6953},
Joseph E. Rodriguez$^{1}$\orcidlink{0000-0001-8812-0565},
Allyson Bieryla$^{3,4}$\orcidlink{0000-0001-6637-5401},
George Zhou$^{4}$\orcidlink{0000-0002-4891-3517},
\newauthor
Theron W. Carmichael$^{5}$\orcidlink{0000-0001-6416-1274},
Steven Giacalone$^{6}$\orcidlink{0000-0002-8965-3969},
Jeffrey D. Crane$^{7}$\orcidlink{0000-0002-5226-787X},
Stephen A. Shectman$^{7}$\orcidlink{0000-0002-8681-6136},
\newauthor
Johanna Teske$^{7,8}$\orcidlink{0009-0008-2801-5040}
\\
$^1$\msu\\
$^2$Department of Astronomy, University of Illinois at Urbana-Champaign, Urbana, IL 61801, USA\\
$^3$\cfa\\
$^4$\usq\\
$^5$Institute for Astronomy, University of Hawai‘i, 2680 Woodlawn Drive, Honolulu, HI 96822, USA\\
$^6$Department of Astronomy, California Institute of Technology, Pasadena, CA 91125, USA\\
$^{7}$The Observatories of the Carnegie Institution for Science, 813 Santa Barbara Street, Pasadena, CA 91101, USA\\
$^{8}$Earth and Planets Laboratory, Carnegie Institution for Science, 5241 Broad Branch Road, NW, Washington, DC 20015, USA\\
}

\date{}

\pubyear{{\the\year{}}}

\begin{document}
\label{firstpage}
\pagerange{\pageref{firstpage}--\pageref{lastpage}}
\maketitle

\begin{abstract}
We present the first measurement of the sky-projected orbital obliquity of a benchmark transiting brown dwarf host, HIP 33609, as a part of the Orbital Architectures of Transiting Massive Exoplanets And Low-mass stars (OATMEAL) survey. HIP 33609 b is a highly eccentric, 68 \mj\ brown dwarf orbiting a 10\,300 K, A-type star with an orbital period of 39 days. Its host star is a known member of the 150 Myr old MELANGE-6 moving group, making it an excellent laboratory for testing sub-stellar evolutionary models. Using in-transit spectra collected by the Planet Finder Spectrograph (PFS) on the Magellan II Clay 6.5 m telescope, we measured a sky-projected orbital obliquity of \thislambda. The mass of the brown dwarf is most consistent with a stellar-like fragmentation formation history followed by a period of migration. Given the high eccentricity ($e=0.557$) but low orbital obliquity of the brown dwarf, we claim that coplanar high eccentricity tidal migration seems to be the most plausible pathway, however, it remains difficult to conclusively rule out other migration mechanisms. The low orbital obliquity for HIP 33609 is consistent with previous measurements of high mass-ratio companions, and bears a striking resemblance to the obliquity distribution of transiting warm Jupiters. We suggest brown dwarfs may follow a dynamically quiescent migration pathway, consistent with them forming in isolated conditions. 
\end{abstract}

\begin{keywords}
techniques:spectroscopic -- brown dwarfs -- binaries: close -- planets and satellites:general
\end{keywords}

\section{Introduction}
Brown dwarfs (BDs) have traditionally been distinguished from giant planets and stars by the presence (and type) of fusion that occurs in their cores. Planets are generally not massive enough to ignite nuclear fusion, whereas stars undergo hydrogen fusion. BDs, which lie between these two regimes in mass, are only massive enough to fuse deuterium in their cores, a heavy isotope of hydrogen composed of one proton and one neutron. Nominally, this translates to a boundary of 13 \mj\ to ignite deuterium fusion \citep{Spiegel2011} and 80 \mj\ to ignite hydrogen fusion \citep{Baraffe2003}. While this fusion-based definition is useful for understanding the underlying physics operating in their interiors, these objects can also be understood based on how they form and evolve. From the perspective of formation, there are generally two pathways: objects that form via core accretion \citep[planet-like;][]{Pollack:1996}, and objects that form via gravitational instability \citep[star-like;][]{Adams1989, Bate2012, Kratter2016}. Objects in the canonical BD regime can theoretically form through either formation channel, but it is not yet clear under what conditions each mechanism dominates.

Although differentiating between these mechanisms remains challenging for any individual BD, a large enough population of objects may show trends that favor one mechanism over the other. Several recent studies have attempted to explore this with different BD populations. \citet{Ma&Ge2014} used the radial velocity (RV) and transit detections at the time to show that brown dwarf architectures support a 42.5 \mj\ transition between planet-like and star-like formation pathways. \citet{Schlaufman2018} used an updated sample to explore host star metallicities, finding evidence for a much lower $\sim$4 \mj\ transition point. \citet{Bowler2020, Bowler2023} showed that directly imaged giant planets on widely separated orbits have very different architectures than their BD counterparts. Specifically, they found that the relative alignment of the companion's orbital plane to its host star's rotation axis (commonly referred to as stellar obliquity) is often much larger for BD companions than giant planets. The authors claim that this trend is supportive of two distinct formation channels since hydrodynamical simulations show that binaries formed via fragmentation can often form with high misalignments \citep{Bate2010, Offner2016}. 

In the era of NASA's Transiting Exoplanet Survey Satellite \citep[\textit{TESS};][]{Ricker:2015} mission, the population of transiting BD systems continued to grow in the wake of recent discoveries \citep[e.g.][]{Grieves2021, Carmichael2022, Henderson2024b, Vowell2025}. With more than 50 transiting BDs now discovered, the population has just grown large enough to begin searching for trends without being biased by a small sample size. One of the most anticipated parameters to explore in search of trends that can be tied to formation is stellar obliquity because of its usefulness in distinguishing the evolutionary pathways of close-in giant planets \citep{Rice2022a} as well as widely-separated massive companions \citep{Bowler2023}. However, only 10 transiting BDs have had their stellar obliquity directly measured so far \citep{Triaud2009, Siverd:2012, Triaud2013, Zhou2019a, Giacalone2024, dosSantos2024, Brady2025, Doyle2025, Rusznak2025, Carmichael2025}. 4 of these systems have eccentric orbits and all 10 of them lie on orbital periods $P<10$ days. Overall, the transiting BDs exhibit a strong preference for alignment among single-star hosts \citep{Rusznak2025}. 

The stellar obliquity of transiting companions is typically measured through the Rossiter-McLaughlin (RM) effect \citep{Rossiter:1924, McLaughlin:1924}, an observed RV shift that occurs during a companion's transit. This observed RV shift is directly tied to the specific transit chord that is traced on the stellar surface by the transiting companion. As the companion transits, the local rotational velocity where the stellar surface is blocked will not contribute to the broadened line profile. The resulting distortion of line shape causes the apparent RV shift. This effect becomes less observable in rapidly rotating stars due to excessive broadening of the star's spectral lines; however, in some cases, the stellar obliquity can still be retrieved with techniques that measure the distortion of the spectral lines themselves rather than the induced RV shift (e.g. Doppler Tomography; \citealt{CollierCameron:2010}, Reloaded Rossiter-McLaughlin; \citealt{Cegla:2016}). Here, we present a new stellar obliquity measurement for a transiting brown dwarf, HIP 33609 b, using the Doppler Tomography\footnote{Doppler Tomography is also sometimes referred to as the Doppler Shadow technique.} technique. 

HIP 33609 b is a benchmark young (150 Myr) transiting brown dwarf in the comoving group MELANGE-6 \citep{Vowell2023}. It orbits a bright, \textit{V} = 7.3 mag host star making it easily accessible for follow-up characterization. It also lies on a 39-day, highly eccentric orbit. This architecture is consistent with a dynamically active past which is theorized to also produce potentially large misalignments \citep{Kozai1962, Lidov1962, Rasio:1996, Chatterjee2008, Wu2011}. These factors make HIP 33609 an ideal candidate to probe for misalignment since it is also tidally detached and therefore would not have had any potential misalignment dampened by tidal interactions. 
This paper is structured as follows: in \S\ref{sec:Observations} we present the archival and in-transit spectroscopic observations of HIP 33609 b used in our analysis. We describe our analysis of these data to measure the stellar obliquity in \S\ref{sec:Analysis}. In \S\ref{sec:Discussion} we discuss the most probable formation and evolutionary pathways for HIP 33609 b given it's newly measured stellar obliquity, and we place this system in context with the population of previously measured giant planet and BD stellar obliquities. Finally, we present our conclusions in \S\ref{sec:Conclusion}.

\section{Observations}
\label{sec:Observations}
We used photometric time-series data, out-of-transit radial velocities, and archival photometry to globally fit the host star and Keplerian orbit of the HIP 33609 system. Most of the following observations were originally collected and analyzed during the confirmation process detailed in \citet{Vowell2023}. This included photometric time-series data from NASA's Transiting Exoplanet Survey Satellite (\textit{TESS}) mission \citep{Ricker:2015}, out-of-transit RVs from the CHIRON spectrograph on the 1.5 m SMARTS telescope at the Cerro Tololo Inter-American Observatory \citep{Tokovinin2013, Paredes2021}, and broadband photometry from Gaia Data Release 3 \citep{GaiaDR3}, 2MASS \citep{Cutri:2003, Skrutskie:2006}, and WISE \citep{Wright:2010, Cutri:2012}. 

Since the original discovery, \textit{TESS} has reobserved HIP 33609 during sectors 61, 87, and 88. Due to the long orbital period ($\sim$ 39 days), transits were only observed in sectors 61 and 88. Unfortunately, the sector 88 transit occurred during the spacecraft's momentum dump causing a period of unstable pointing which resulted in poor data quality. We therefore were only able to add the sector 61 data to the data from \citet{Vowell2023} for our global analysis. The archival photometric and spectroscopic data, and the new transit from sector 61 were collected and reduced according to \citet{Vowell2023}.

We also observed a transit of HIP 33609 b with the Planet Finder Spectrograph (PFS) on the Magellan II Clay 6.5 m telescope \citep{Crane2006, Crane2008, Crane2010} on the night of 2023-02-05 UTC to measure the sky-projected stellar obliquity of the system. Due to the long transit duration of HIP 33609 b ($\sim 6.5$ hours), we were only able to observe the first half of the transit before the target fell too low in the sky to continue observing. The observations were taken using the $0.3^" \times 2.5^"$ slit with a resolving power of $R=127,000$. We did not use the iodine cell. A total of 43 exposures were taken with an exposure time of 300 s. We reduced these spectra according to \citep{Butler1996}.

In order to measure the projected stellar obliquity, we utilized Doppler Tomography \citep{CollierCameron:2010}. This technique leverages the fact that a transiting companion will block out different parts of the host star's surface, which possess their own local rotational velocity. By tracking the companion's shadow projected on the host star's rotationally broadened line-profile, we were able to reconstruct the precise transit chord as a function of the local rotational velocity. To do this, we first derived the line broadening profile with a least-squares deconvolution between the data and a synthetic spectrum based on the ATLAS9 model atmospheres \citep{Castelli2003} which best matched the parameters of the host star. We calculated the local rotational velocity of the portion of the stellar surface being blocked out by subtracting the median combined line profile from the derived line profile. For a more in depth description of this process see \citet{Zhou2019a}. We then fit this Doppler shadow to a model in order to constrain the projected orbital obliquity of the system, see \S\ref{sec:Analysis}

\section{Analysis and Results}
\label{sec:Analysis}

\begin{table*}
\begin{center}
\caption{Median Values and 68\% Confidence Intervals for Fitted Stellar and Companion Parameters. \label{tab:toi588}}
\end{center}
\begin{tabular}{llc}
\hline
\hline
Parameter & Description & Value \\
\hline
\multicolumn{3}{l}{\textbf{Priors}} \\
$\pi$ & Gaia parallax (mas) & $\mathcal{G}$[6.4871, 0.0492] \\
$[{\rm Fe/H}]$ & Metallicity (dex) & $\mathcal{G}$[0.0, 0.5] \\
$A_V$ & V-band extinction (mag) & $\mathcal{U}$[0, 0.0358] \\
$D_T$ & Dilution in \textit{TESS} & $\mathcal{G}$[0, 0.00162] \\
\hline
\multicolumn{3}{l}{\textbf{Stellar Parameters}} \\
$M_*$ & Mass ($M_\odot$) & $2.375^{+0.098}_{-0.093}$ \\
$R_*$ & Radius ($R_\odot$) & $1.865^{+0.080}_{-0.075}$ \\
$L_*$ & Luminosity ($L_\odot$) & $35.6^{+10.0}_{-6.6}$ \\
$F_{Bol}$ & Bolometric Flux (cgs) & $4.78^{+1.4}_{-0.89} \times 10^{-8}$ \\
$\rho_*$ & Density (cgs) & $0.515^{+0.070}_{-0.061}$ \\
$\log{g}$ & Surface gravity (cgs) & $4.272^{+0.039}_{-0.037}$ \\
$T_{\rm eff}$ & Effective temperature (K) & $10320^{+800}_{-620}$ \\
$[{\rm Fe/H}]$ & Metallicity (dex) & $0.01^{+0.18}_{-0.20}$ \\
$[{\rm Fe/H}]_{0}$ & Initial Metallicity & $0.03^{+0.17}_{-0.19}$ \\
Age & Age (Gyr) & $0.153^{+0.024}_{-0.025}$ \\
EEP & Equal Evolutionary Phase & $313.9^{+7.2}_{-7.3}$ \\
$A_V$ & V-band extinction (mag) & $0.125^{+0.078}_{-0.080}$ \\
$d$ & Distance (pc) & $154.3 \pm 1.2$ \\

\hline
\multicolumn{3}{l}{\textbf{Companion Parameters}} \\
$P$ & Period (days) & $39.4718115 \pm 0.0000072$ \\
$R_P$ & Radius ($R_{\rm J}$) & $1.581^{+0.068}_{-0.064}$ \\
$M_P$ & Mass ($M_{\rm J}$) & $67.9^{+7.3}_{-7.2}$ \\
$T_C$ & Time of conjunction (BJD$_{\rm TDB}$) & $2459231.75945 \pm 0.00013$ \\
$T_0$ & Optimal Conjunction Time (BJD$_{\rm TDB}$) & $2459271.23166 \pm 0.00010$ \\
$a$ & Semi-major axis (AU) & $0.3054^{+0.0041}_{-0.0040}$ \\
$i$ & Inclination (degrees) & $89.03 \pm 0.12$ \\
$e$ & Eccentricity & $0.557^{+0.029}_{-0.030}$ \\
$\omega_*$ & Argument of periastron (degrees) & $167.4^{+5.1}_{-5.3}$ \\
$T_{\rm eq}$ & Equilibrium temperature (K) & $1230^{+75}_{-57}$ \\
$K$ & RV semi-amplitude (m/s) & $2700 \pm 290$ \\
$R_P/R_*$ & Planet radius in stellar radii & $0.08709^{+0.00031}_{-0.00032}$ \\
$a/R_*$ & Semi-major axis in stellar radii & $35.2^{+1.5}_{-1.4}$ \\
$\delta$ & $(R_P/R_*)^2$ & $0.007584 \pm 0.000055$ \\
$\delta_{\rm TESS}$ & Transit depth in TESS & $0.00874 \pm 0.00023$ \\
$\tau$ & Ingress/Egress duration (days) & $0.02481^{+0.00070}_{-0.00071}$ \\
$T_{14}$ & Total transit duration (days) & $0.27057 \pm 0.00061$ \\
$b$ & Transit impact parameter & $0.369^{+0.029}_{-0.034}$ \\
$b_S$ & Predicted eclipse impact parameter & $0.468^{+0.057}_{-0.054}$ \\
$\tau_S$ & Predicted ingress/egress eclipse duration (days) & $0.0333^{+0.0045}_{-0.0038}$ \\
$T_{S,14}$ & Predicted total eclipse duration (days) & $0.331^{+0.025}_{-0.024}$ \\
$\delta_{S,2.5\mu m}$ & Predicted eclipse depth at 2.5$\mu$m (ppm) & $53.2^{+10.0}_{-7.5}$ \\
$\delta_{S,5.0\mu m}$ & Predicted eclipse depth at 5.0$\mu$m (ppm) & $260^{+18}_{-15}$ \\
$\delta_{S,7.5\mu m}$ & Predicted eclipse depth at 7.5$\mu$m (ppm) & $413^{+15}_{-14}$ \\
$\rho_P$ & Planet density (cgs) & $21.3^{+3.7}_{-3.2}$ \\
$\log g_P$ & Planet surface gravity (cgs) & $4.828^{+0.056}_{-0.059}$ \\

\hline
\end{tabular}
\end{table*}

We used \exofast\ \citep{Eastman:2019} to update the global fit performed in \citet{Vowell2023}. Generally, we followed the same techniques described in \citet{Vowell2023} except that we chose not to include the ground-based photometry collected by the \textit{TESS} Follow-up Observing Program \citep[TFOP;][]{Collins:2018}. Generally, these data are most useful during the confirmation process where they can rule out various false positive scenarios. They also often increase the precision of the ephemeris by extending the observational baseline, since the follow-up light curves are often collected after the most recent \textit{TESS} observations. In this case however, the substellar nature of the transit signal has already been confirmed, and \textit{TESS} has since revisited this field providing more data after the ground-based, follow-up data were collected. Hence, these follow-up data provide little value to the fit while simultaneously adding computational complexity, and we therefore choose not to include them in this analysis.

To briefly summarize the fitting process, we used archival photometry from Gaia Data Release 3 \citep[DR3][]{GaiaDR3}, 2MASS \citep{Cutri:2003, Skrutskie:2006}, and WISE \citep{Wright:2010, Cutri:2012}, in combination with the \textit{TESS} light curves and radial velocities described in \citep{Vowell2023}. Using these data, we fit the host star using a combination of a Spectral Energy Distribution (SED) model and MESA Isochrones and Stellar Tracks (MIST). We also simultaneously fit the substellar companion by fitting the light curve and radial velocities to a Keplerian model. We placed the same Gaussian priors as \citet{Vowell2023} on the Gaia parallax, [Fe/H], and dilution in the \textit{TESS} band. We also used the same uniform prior on the \textit{V}-band extinction. We initialized the fit by placing starting points for each parameter at the median value reported in \citep{Vowell2023}. The posteriors resulting from this fit (see Table \ref{tab:toi588}) agree with the posteriors in \citet{Vowell2023} to 1$\sigma$. 

We then perform two separate fits to the Doppler shadow. For the first fit, we incorporated the posteriors from our \exofast\ global fit as priors on the Doppler shadow fit. Specifically, we applied Gaussian priors using the posteriors for mid-transit time ($T_C$), orbital period ($P$), $R_P/R_*$, impact parameter $b$, eccentricity ($e$) and argument of periastron ($\omega_*$). For the second fit, we model each of these parameters together, incorporating the \textit{TESS} lightcurves, CHIRON RVs and PFS spectra.
Both fits incorporated the projected rotational velocity of the host star ($v\sin{i_\star}$) from \citet{Vowell2023} as a prior. We ran both fits according to the procedures outlined in \citet{Dong2022}. To briefly summarize, we fit the Doppler shadow to a model created by calculating the planet's position on the stellar disk, as well as the local rotational velocity on stellar surface being blocked by the shadow. Then, we convolved this with a Gaussian velocity profile representing the resolution of the spectrograph and macroturbulence of the star. This normalized planetary velocity was then compared to the observed Doppler shadow to calculate the likelihood. 

We modeled the Doppler shadow using the \texttt{PyMC} package with the No-U-Turn Sampler (NUTS) for posterior inference. We ran four independent Markov chains, each with 5000 tuning steps followed by 3000 draws. All model parameters achieved $\hat{R} < 1.01$, indicating good convergence of the posterior distributions. The posteriors from both these fits as well as the parameters they shared with the global \exofast\ fit were all consistent with each other within 1$\sigma$. We adopted the results from the second fit, which simultaneously models all parameters, as it incorporates the full data set. The resulting stellar obliquity from that fit was \thislambda.

\begin{figure*}
    \centering
    \includegraphics[width=\linewidth]{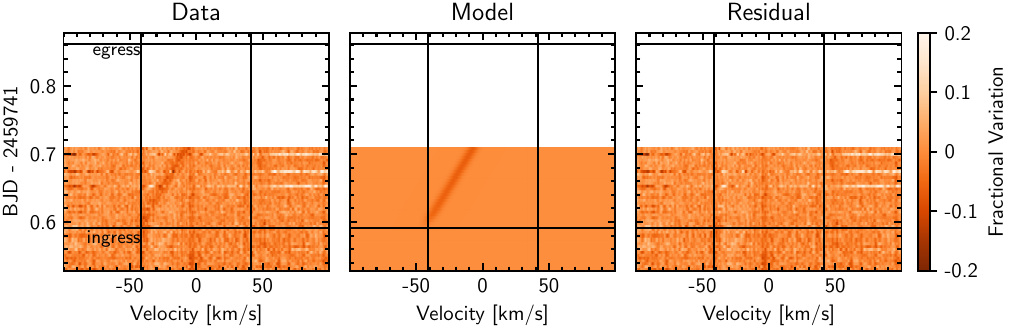}
    \caption{Doppler shadow analysis of HIP 33609~b. The left, middle, and right panels show the reduced data, best-fit model, and residuals (data minus model), respectively. The horizontal lines indicate the times of ingress and egress, while the vertical lines mark the projected stellar rotational velocity limits, $\pm v \sin i$. The color scale represents the fractional variation in flux. \label{fig:dt}}
\end{figure*}

\section{Discussion}
\label{sec:Discussion}

\begin{figure*}
\centering 
\includegraphics[width=\linewidth]{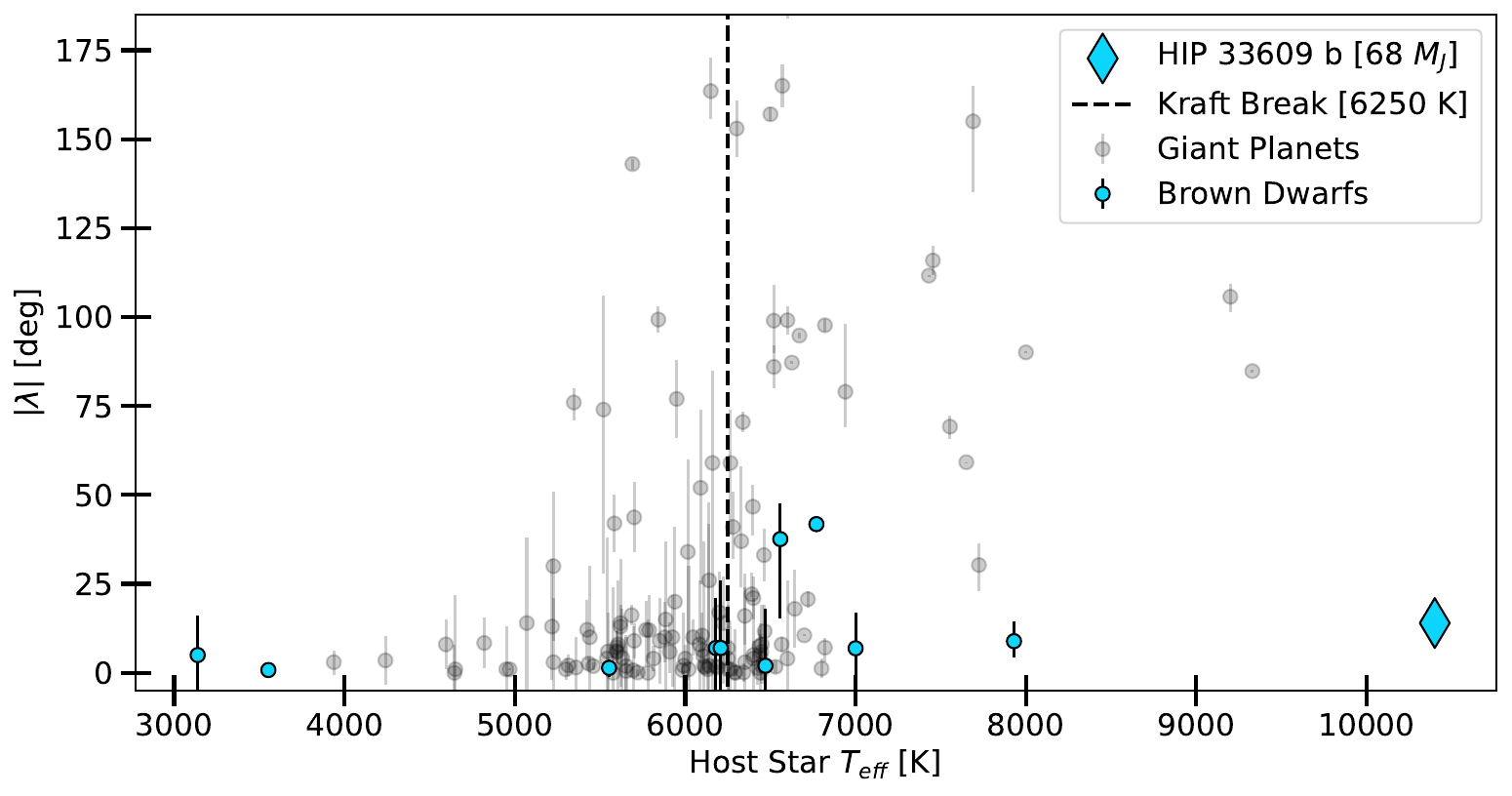}
\caption{The sky-projected stellar obliquity vs. host star $T_{\rm eff}$ for transiting companions ranging $0.7<M_b<80$ $M_{\rm J}$. The vertical dashed line at $6250$ K marks the Kraft break \citep{Kraft:1967}. Giant planets planets (gray) transition from low to high stellar obliquity at the Kraft break, whereas BDs (blue) exhibit low stellar obliquity regardless of their host star's effective temperature. HIP 33609 b stands out as the hottest stellar obliquity measurement to date and is denoted by the blue diamond. All data for this figure were retrieved using TEPCat \citep{Southworth2011}.}
\label{fig:tefflambda}
\end{figure*}

HIP 33609 b is now the 11th transiting BD with a measured sky-projected stellar obliquity (see Figure \ref{fig:tefflambda}), and within this population, it has both the longest orbital period ($\sim39$ days) and the hottest host star ($T_{\rm eff} = 10\,300$ K). These factors, combined with the young age ($150 \pm 25$ Myr) constrained via cluster membership means that HIP 33609 b has likely not undergone any significant tidal realignment. Hot stars ($T_{\rm eff}>6250$) have been shown to be less efficient at realigning transiting hot Jupiters, even on Gyr timescales \citep{Rice2022a}, and HIP 33609 b is several factors further from its host star ($a/R_* = 35.2^{+1.5}_{-1.4}$) than most hot Jupiters. The timescale for tidal realignment is a strong function $a/R_*$, see Equation 3 in \citet{Giacalone2024} adapted from \citet{Zahn1975, Zahn1977, Albrecht:2012}, so HIP 33609 b's architecture and age can rule out any significant tidal realignment. Given that lack of tidal influence, we examine the formation and evolution of the BD companion below. We also place this system in context with previously measured giant planet and BD stellar obliquities, noting the similarities between the BD and warm Jupiter stellar obliquity distributions.

\subsection{HIP 33609 b Evolutionary History}
\label{sec:evohistory}
While it remains challenging to discern between planet-like formation (core accretion) and stellar-like formation (disk fragmentation) for any individual transiting BD, the measurement of HIP 33609 b's sky-projected orbital obliquity, provides yet another piece of the puzzle for understanding its evolutionary history. By combining this measurement with the other known properties of the HIP 33609 system, we can hone in a probable evolutionary pathway. First, with a mass of 68 \mj, the core accretion formation mechanism is unlikely \citep{Mordasini2009A}. This is mainly due to the limited lifetime of the circumstellar disk around FGK type stars which truncates the runaway gas accretion phase, inhibiting planet growth past $\sim 20$ \mj\ \citep{Mordasini2012}. In the case of HIP 33609, the disk lifetime is expected to be even shorter than typical transiting BD hosts due to its more massive host star which would further inhibit growth via core accretion \citep{Richert2018}. This fact alone makes disk fragmentation the more likely culprit behind the formation of HIP 33609 b. 

However, both disk fragmentation and core accretion struggle to directly reproduce the current architecture of the HIP 33609 system at birth. In the case of core accretion, the formation of massive planets becomes inefficient when the semi-major axis a $<$ 1 AU \citep{Dawson:2018}. Similarly, stellar companions on comparable orbits are also rare at birth \citep{Moe2017}. Therefore, both scenarios require migration to explain the current architecture. The high eccentricity of the BD (e = 0.56) offers some clues into its migration history. For example, high eccentricity tidal migration triggered by a nearby massive companion could explain the highly eccentric orbit. These include Kozai-Lidov oscillations \citep{Kozai1962, Lidov1962}, planet-planet scattering \citep{Rasio:1996, Chatterjee2008}, and secular interactions \citep{Petrovich2015, Wu2011} which are often invoked to explain the origins of short-period giant planets \citep{Dawson:2018}. Recent studies have shown that interactions between the protoplanetary disk and a massive planet or BD can also produce highly eccentric orbits \citep{Li2023, Romanova2023}. In this scenario, an eccentric HIP 33609 b would be massive enough to also torque the protoplanetary disk, which could also induce migration. 

Stellar mass companions with orbital periods P $<$ 100 days are also thought to migrate through the same dynamical and secular interactions as the hot Jupiters \citep{Moe2017}. In other words, the mechanism of HIP 33609 b's migration should be agnostic of its formation mechanism. Distinguishing between the various migration pathways described above requires additional information. For example, the interactions that require the presence of some other companion capable of driving up the eccentricity can favored when an outer companion is detected. However, we did not detect any nearby companions in the high resolution imaging, nor any additional transits in the \textit{TESS} photometry. The Gaia Re-normalized Unit Weight Error (RUWE) is low (RUWE = 0.97), signifying that the Gaia astrometric solution is well-described by a single component. The radial velocities also show no sign of a longer term trend over the 3 observational epochs spanning $\sim$2 years that may indicate outer bodies. While we are able to rule out additional short-period, large transiting companions, and widely-separated, bright companions, there is still ample parameter space where a companion of at least M-dwarf mass could reside and remain undetected by the available data. Such a companion would certainly be capable of inducing the migration of HIP 33609 b. Thus high eccentricity migration is still plausible given the ample parameter space to hide a massive companion.

Knowledge of the relative alignment of the companion's orbit can also help illuminate migration history since, each of the dynamical interactions discussed above predict unique stellar obliquity distributions. Therefore, we can attempt to rule out mechanisms that are inconsistent with HIP 33609 b's measured orbital obliquity. Kozai-Lidov predicts a bimodal distribution with peaks at 60\degree\ and 120\degree\ \citep{Fabrycky:2007} whereas scattering results in a more uniform distribution of stellar obliquities \citep{Chatterjee2008}. Secular chaos tends to induce moderate, prograde stellar obliquities \citep{Wu2011}, and coplanar high eccentricity migration tends to keeps stellar obliquity low \citep{Petrovich2015}. The companion-disk interactions described in \citet{Li2023, Romanova2023} are restricted to the plane of the disk. We measured a sky-projected orbital obliquity of \thislambda, and note that it is unlikely that HIP 33609 b could have undergone significant tidal realignment. Given the system's young age ($150$ Myr), relatively wide separation ($a/R_* = 35.2$), and host star effective temperature ($T_{\rm eff}=10,300$) it would not have had enough time to realign under equilibrium-tide theory \citep{Winn:2010b} or via resonance locking \citep{Zanazzi2024}. Thus, HIP 33609 b's stellar obliquity appears most consistent with coplanar high eccentricity migration, however, it's difficult to completely rule out other migration pathways. Detecting another massive companion in the system would provide strong evidence in favor of a dynamically hot migration mechanism, and indeed the multiplicity frequency for HIP 33609-like stars with late A/early B spectral types is roughly 85\% \citep{Moe2017}. While \textit{TESS}\ will not reobserve this system in year 8, it may be reobserved in a potential 3rd extended mission allowing for the possibility to detect a transit from an outer long-period companion. Similarly, continuous RV monitoring of the system may uncover a long period trend indicative of another massive body as well.

\subsection{Evidence for Quiescent Transiting BD Migration}
\label{sec:populations}
\begin{figure*}
\centering 
\includegraphics[width=\linewidth]{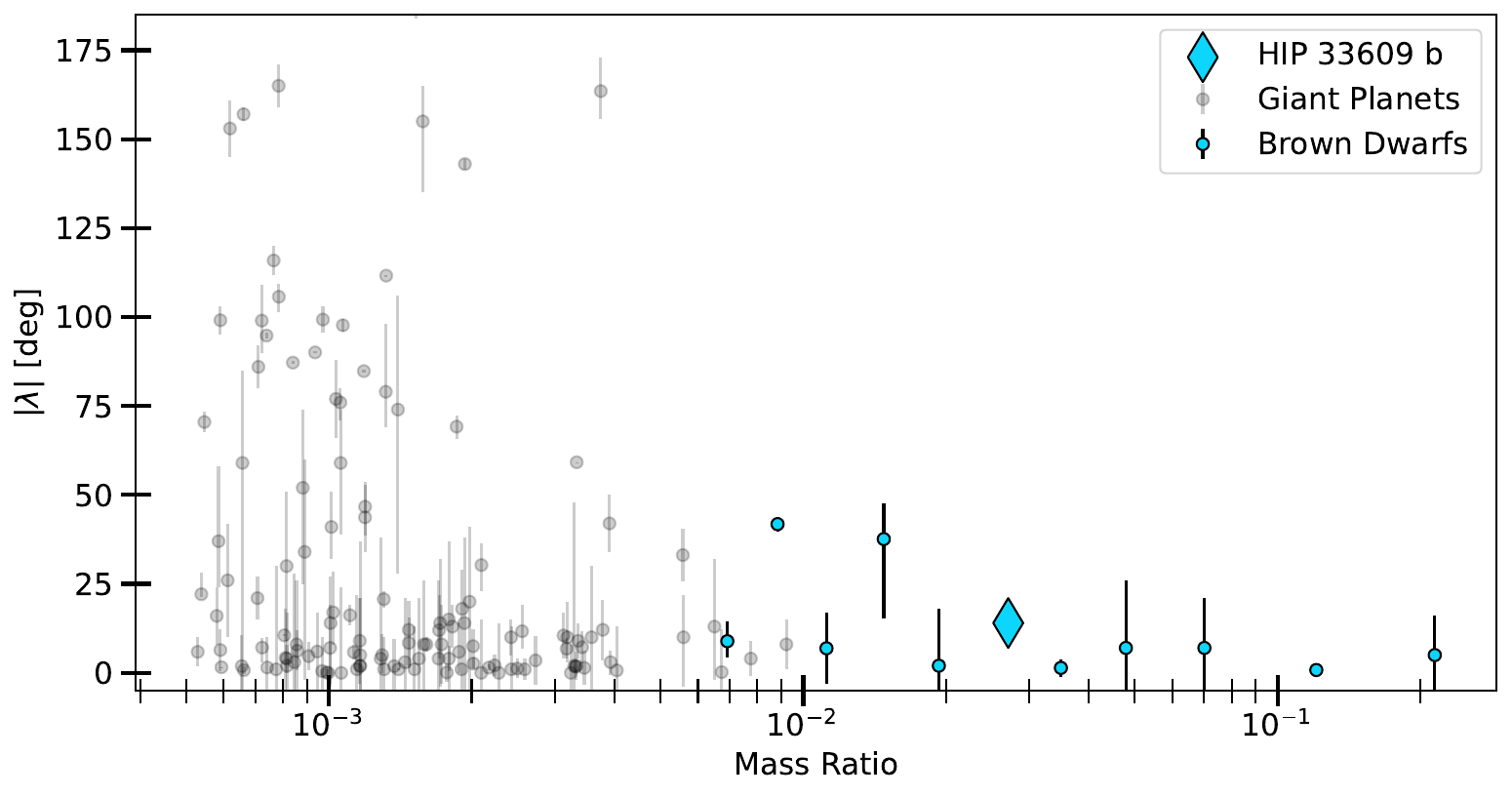}
\caption{The sky-projected stellar obliquity vs. the mass ratio of the system for transiting companions ranging $0.7<M_b<80$ $M_{\rm J}$. Giant planets are shown as gray circles whereas BDs are blue circles. Our new measurement for HIP 33609 b is highlighted as the blue diamond. HIP 33609 b's stellar obliquity is consistent with the \citet{Rusznak2025} framework wherein high mass-ratio companions form isolated with low stellar obliquities.}
\label{fig:qlambda}
\end{figure*}

HIP 33609 b is now the 11th transiting BD with a measured stellar obliquity. It also orbits the hottest host star for which this measurement has been made. This measurement continues the trend presented \citet{Giacalone2024, Doyle2025, Rusznak2025}, which noted that BD companions tend to have significantly lower stellar obliquities than their less massive counterparts (e.g. hot Jupiters), regardless of their host's effective temperature. This is initially perplexing, since transiting BDs are expected to have migrated to their present-day locations under the same mechanisms as hot Jupiters. This is because migration is generally agnostic to the mass of the system, with the same phenomena applying for giant planets up to stellar binaries \citep{Moe2017, Dawson:2018}. The hot Jupiter orbital obliquity distribution, which transitions from low stellar obliquity to high at the Kraft break \citep[$\sim6250$ $K$][]{Kraft:1967}. This has be explained by the fact that hot stars cannot efficiently realign their companions due to their lack of a convective envelope. Under equilibrium-tide theory, realignment results from the interaction between the orbit of the companion and the convective zone of the star \citep{Winn:2010b, Rice2022a}. The combined distributions of hot Jupiter stellar obliquities and eccentricities have been cited as evidence of high-eccentricity migration being the dominant migration mechanism \citep{Rice2022a}. If the transiting BDs migrate predominately by the same mechanisms, then we might expect them to have similar architectures, which vitally includes their orbital obliquities. However, transiting BDs seem to be significantly more aligned with 9 of the 11 total obliquity measurements being consistent with 0 to 3$\sigma$. This is suggestive that their migration pathways may be distinct, see Figure \ref{fig:qlambda}.

One possible explanation is presented in \citet{Rusznak2025} where they claim that low mass-ratio companions ($M_P/M_* < 2 \times 10^{-3}$) may form in compact multiplanet systems where dynamical instabilities capable of driving misalignment are common. Higher mass-ratio companions, on the other hand, may form in more isolated environments that lack massive nearby companions capable of perturbing the growing BD onto an inclined orbit. Therefore, the companion remains in the disk where it can continue to accrete and grow undisturbed. Indeed, the distribution of giant planet and brown dwarf stellar obliquities is consistent with this framework, including our new measurement for HIP 33609 b. However, this mechanism only explains the primordial alignment of these systems, not their subsequent migration, which, as noted above in \S\ref{sec:evohistory}, is required to observe them in their current configurations. Hence, it remains an open question to explain the orbital architectures of transiting BDs since they still need to migrate close to their host stars, but without exciting significant orbital obliquities. Superficially, this appears similar to recent trends emerging in the warm Jupiter population.

\citet{Rice2022b} and \citet{Wang2024} showed that the warm Jupiters around single stars are more aligned than the hot Jupiter population regardless of their host star's effective temperature. Since the warm Jupiters orbit farther and are tidally detached, they should not undergo significant tidal realignment. Both studies also note evidence for primordial alignment among giant planets, suggesting that hot Jupiters migrate through some dynamically violent process that drives high stellar obliquity, whereas warm Jupiters may migrate more quiescently and, therefore, never become misaligned. This framework is also consistent with studies showing that warm Jupiters tend to have nearby companions significantly more frequently than hot Jupiters \citep{Huang:2016}. Even more evidence that warm Jupiters seem to migrate quiescently, since any dynamically hot migration mechanism would have ejected nearby companions \citep{Wu2023}. 

\begin{figure*}
\centering 
\includegraphics[width=\linewidth]{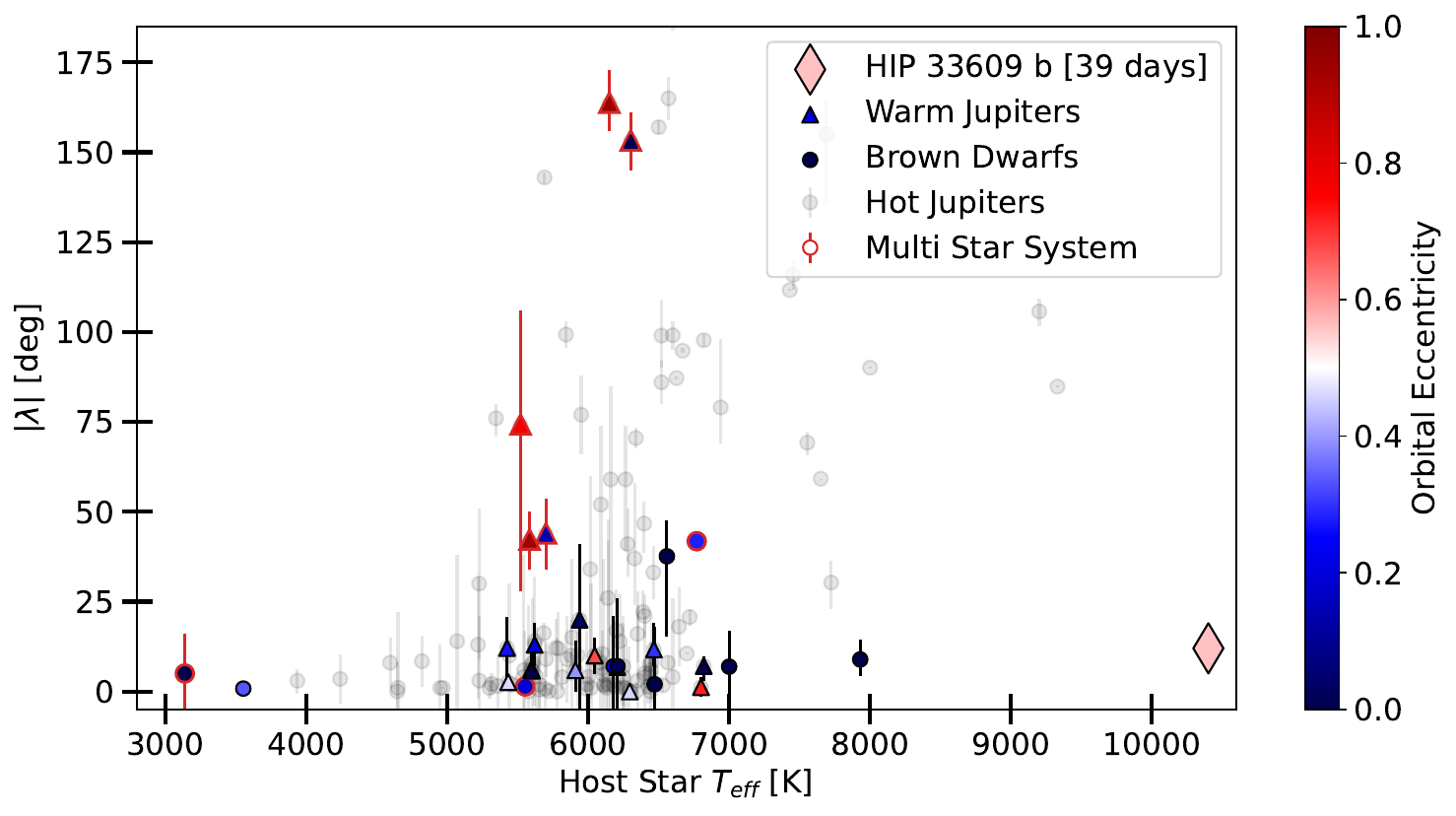}
\caption{The sky-projected stellar obliquity vs. host star $T_{\rm eff}$ for transiting companions ranging $0.7<M_b<80$ $M_{\rm J}$. Warm Jupiters are plotted in color as triangles and BDs as circles. Hot Jupiters are the gray background points. Warm Jupiters and BDs are colored based on their eccentricities. Systems with a detected outer stellar companion are highlighted with a red outline. Warm Jupiters and BDs orbiting single stars exhibit low stellar obliquities regardless of their host star's effective temperature and orbital eccentricity suggesting a potentially shared, quiescent migration pathway (see \S\ref{sec:populations}).}
\label{fig:WJ-BD}
\end{figure*}

It is intriguing then that the BD obliquity distribution seems to more closely resemble the warm Jupiters (see Figure \ref{fig:WJ-BD}) given their relatively low stellar obliquities. While it is true that 6 warm Jupiters or BDs show large misalignments, they are almost entirely confined to systems with a detected outer stellar companion, which can explain their architectures. We restrict our discussion on the obliquity distribution resemblance between warm Jupiters and BDs to single-star systems to maintain consistency with the claims of \citet{Rice2022b} and \citet{Wang2024}. Both studies noted that stellar companions have been shown to strongly influence stellar obliquity, often leading to large misalignments. 

One potential interpretation of the similarity between warm Jupiter and transiting BD stellar obliquities would be that they undergo similar migration mechanisms. This would suggest that the transiting BDs migrate quiescently to their current, short-period orbits in order to retain their primordially low stellar obliquities. This interpretation is consistent with the hypothesis of \citet{Rusznak2025} wherein high mass-ratio companions form isolated due to the fact that isolated formation suggests no massive nearby bodies around to trigger a dynamical instability in the first place. If this is true and transiting BDs do indeed migrate quiescently (like warm Jupiters), then it should follow that they would also not disrupt the orbits of nearby smaller planets as the hot Jupiters have done \citep{Huang:2016}. Therefore, one way to provide more evidence for quiescent migration would be to systematically search the current transiting BD population for nearby small planets to understand the occurrence rate of nearby planetary companions to transiting BDs. 

\section{Conclusion}
\label{sec:Conclusion}
We report the sky-projected orbital obliquity of \thislambda for the transiting BD HIP 33609 b. This marks the 11th transiting BD system for which this measurement has been made. It is also the hottest transiting BD host and longest BD orbital period for which a stellar obliquity measurement has been successful. Equipped with this measurement, and therefore a more complete understanding of the HIP 33609 system, we argued that a stellar-like fragmentation mechanism is the most likely formation channel followed by subsequent high eccentricity tidal migration. Ruling out any specific migration mechanism remains difficult, however, our stellar obliquity measurement appears most consistent with a coplanar high eccentricity migration mechanism. We found that our measurement continues the ongoing trend of low obliquity among high mass-ratio companions, and note the similarity this distribution to the stellar obliquity measurements of warm Jupiters. We claimed that this similarity may be indicative of a shared migration history between the two populations which would imply that transiting BDs migrate relatively quiescently to their current day orbits. This idea is consistent with the results presented in \citet{Rusznak2025}, which claimed that transiting BDs form isolated in the circumstellar disk. Finally, we propose future work to constrain the occurrence rate of nearby small planets to transiting BDs in order to further explore these systems' resemblance to the warm Jupiters. 

\section*{Acknowledgements}
This paper includes data collected by the \textit{TESS} mission that are publicly available from the Mikulski Archive for Space Telescopes \citep[MAST;][]{TICdoi}. Funding for the \textit{TESS} mission is provided by NASA's Science Mission Directorate. We acknowledge the use of public \textit{TESS} data from pipelines at the \textit{TESS} Science Office and at the \textit{TESS} Science Processing Operations Center. Resources supporting this work were provided by the NASA High-End Computing (HEC) Program through the NASA Advanced Supercomputing (NAS) Division at Ames Research Center for the production of the SPOC data products. This research has made use of the NASA Exoplanet Archive and the Exoplanet Follow-up Observation Program (ExoFOP; DOI: 10.26134/ExoFOP5) website, which is operated by the California Institute of Technology, under contract with the National Aeronautics and Space Administration under the Exoplanet Exploration Program. NV is supported by the NASA FINESST program. TWC is supported by an NSF MPS-Ascend Postdoctoral Fellowship under award 2316566.


\section*{Data Availability}
The \textit{TESS} light curves described in \S\ref{sec:Observations} are publicly available through MAST\footnote{\url{https://archive.stsci.edu/}}. The out-of-transit radial velocities described in \S\ref{sec:Observations} are available in \citet{Vowell2023}.
The in-transit-spectra described in \S\ref{sec:Observations} are available upon reasonable request to the corresponding author. The stellar obliquity data used in this paper for Figures \ref{fig:tefflambda} - \ref{fig:WJ-BD} are publicly available via TEPCat\footnote{\url{https://www.astro.keele.ac.uk/jkt/tepcat/}}.



\bibliographystyle{mnras}
\bibliography{refs} 


\bsp	
\label{lastpage}
\end{document}